\documentclass[12pt,a4paper]{article}
\usepackage[]{epsfig}	       
\begin{document}
 
 \thispagestyle{empty}
 
 \title{Harmonic polynomials  for expanding the
  fluctuations of the Cosmic Microwave Background: The Poincar\'{e} 
  and the 3-sphere model.}
 \author{Peter Kramer, Institut f\"{u}r Theoretische Physik der 
 Universit\"{a}t\\
 72076 T\"{u}bingen, Germany}
 \maketitle

\section{Abstract.}
Fluctuations of the  Cosmic Microwave Background CMB are observed by the WMAP.  
When expanded   into  the harmonic eigenmodes of the  
space part of a cosmological model, they provide
insight into the large-scale topology of space. 
All harmonic polynomials on the multiply connected
dodecahedral Poincar\'{e} space are constructed.  
Strong and specific selection rules are given by comparing 
the polynomials to those on the 3-sphere,  
its simply connected cover.

\section{Motivation from cosmic topology.} 
The global topology of 3-space is not fixed by Einstein general relativity, 
since this  is formulated in terms of local differential equations. 
Einsteins first static cosmological models used for the space-part
of the universe 
a simply connected sphere $S^3$. With present-day cosmological information
it becomes possible to test multiply-connected topologies for 
the space-part of the universe.
J-P Luminet et al. \cite{LU} 2003 and  J Weeks \cite{WE} 2004 propose 
to explore the topology of 3-space 
from temperature fluctuations 
of the cosmic microwave background (CMB). These fluctuations are measured 
by the Wilkinson Microwave Anisotropy Probe (WMAP) with very high precision.

A way to test the topology is to expand the temperature fluctuations of the CMB  
into  harmonic polynomials of the chosen topological 3-manifold, 
for example the Poincar\'{e} dodecahedral 
manifold ${\cal P}$. The topology will be verified if the harmonic polynomials
of the manifold suffice to expand these fluctuations. 
The simply connected space parts of cosmological models are
the 3-sphere $S^3$ for positive, and the hyperbolic space $H^3$ for
negative curvature. The manifold  ${\cal P}$ has $S^3$ as its 
universal cover and so has positive curvature.

As the backbone for  such an expansion, in what follows we   
characterize an orthogonal 
basis of harmonic polynomials on ${\cal P}$. The details of 
this 
characterization by a novel operator were given in
P. Kramer \cite{KR} and gr-qc/0410094. 
Here we describe   the main steps in group and 
representation theory  for the analysis.

The present construction  allows to compare  the harmonic polynomials of both manifolds. 
By restriction of the topology from $S^3$ to ${\cal P}$ 
we derive  strong and specific selection rules 
for the harmonic polynomials.

\section{General notions from topology.}

For general notions of topology we refer to the classical monograph by Seifert and Threlfall \cite{SE1}, 1934.
The topology of a manifold ${\cal M}$ is characterized by its 
homotopy group $\pi_1({\cal M})$ \cite{SE1} pp. 149-80.
This group operates on ${\cal M}$ by loop composition. If the manifold is 
multiply connected, the homotopy group is non-trivial. Associated to ${\cal M}$
is its simply-connected universal cover $\tilde{{\cal M}}$. The topological
manifold ${\cal M}$ appears on its universal cover $\tilde{{\cal M}}$
in the form of a tiling into copies of  ${\cal M}$. There is a group of deck
transformations $deck(\tilde{{\cal M}})$ \cite{SE1} pp. 195-97.
It  acts fixpoint-free on 
$\tilde{{\cal M}}$ and produces the tiling. This group is 
isomorphic to the homotopy group, 
\begin{equation}
\label{h1}
deck(\tilde{{\cal M}}) \sim \pi_1({\cal M}).
\end{equation}
These relations allow to work out the topology on the universal cover 
and to view the topological manifold ${\cal M}$ as the quotient space
\begin{equation}
\label{h1a}
{\cal M} = \tilde{{\cal M}}/deck(\tilde{{\cal M}}).
\end{equation}

\section{Topology of the Poincar\'{e} dodecahedral\\ 3-manifold ${\cal P}$.} 

H Poincar\'{e} in 1895 introduced the  dodecahedral manifold ${\cal P}$.
C Weber and H Seifert in 1933 \cite{SE2} gave a gluing prescription for ${\cal P}$: 
Glue all pairs of opposite faces of a  spherical dodecahedron, after rotation by $\pi/5$,
to get the topological manifold ${\cal P}$. 
H Seifert and W Threlfall \cite{SE1} pp. 214-19 derived from the 
gluing prescription  the homotopy 
group $\pi_1({\bf {\cal P}})$.
Their proof requires non-trivial steps in combinatorial group theory
to transform from  the original gluing generators and their
relations to new ones. These new generators are then 
shown to belong to the binary icosahedral group ${\cal H}_3$
of order $|{\cal H}_3|=120$ without reflections,
\begin{equation}
\label{h2}
\pi_1({\bf {\cal P}})\sim {\cal H}_3<SU(2,C). 
\end{equation}
The group  ${\cal H}_3$  consists of the preimages in $SU(2)$ of all 
the rotations of the familiar icosahedral group, which is isomorphic to the 
alternating group $A_5$ of five objects. 

We now wish to view the topology on the universal cover.
The Poincar\'{e} dodecahedral manifold 
${\bf {\cal P}}$ has as universal cover  the 3-sphere,
$\tilde{{\bf \cal P}}=S^3$ 
of constant positive curvature $\kappa=+1$.
By the isomorphism eq. \ref{h1}, there is an action
\begin{equation}
\label{h3}  
deck(\tilde{\cal P}) \times \tilde{\cal P } \rightarrow \tilde{\cal P}
\end{equation}
such that  $\tilde{\cal P}$ is tiled by images under ${\cal H}_3$ 
of a prototile ${\cal P}$, presently  a spherical dodecahedron. Conversely from eq. \ref{h1a}, the Poincar\'{e} manifold  
may be taken as the quotient  
\begin{equation}
\label{h4}
{\cal P}=S^3/deck(\tilde{{\cal P}})=S^3/{\cal H}_3.
\end{equation}
We shall work on the universal cover $S^3$.

\section{Coordinates and Lie group actions on $S^3$.}

The sphere  $S^3$ itself is a homogeneous space,
\begin{equation}
\label{h5}
S^3:= SO(4,R)/SO(3,R). 
\end{equation}
Moreover $S^3$ as a manifold is in one-to-one correspondence to  $SU(2)$,
so that the pair $(z_1,z_2)$ of complex numbers may serve as its coordinates, 
\begin{equation}
\label{h6}
u:=
 \left[
\begin{array}{ll}
z_1& z_2\\
-\overline{z}_2&\overline{z}_1
\end{array}
\right],\; z_1\overline{z}_1+z_2\overline{z}_2=1.
\end{equation}

In the coordinates eq. \ref{h6}, $S^3$ admits the following left and right 
actions:
\begin{eqnarray}
\label{h7}
u \in S^3, (g_l, g_r) &\in&  SU(2):
\\ \nonumber
 ((g_l,g_r)\times  u) &=& g_l^{-1}ug_r.
\end{eqnarray}
The left and right actions $(g_l, e), (e,g_r)$ commute. Moreover the full group 
$SO(4,R)$ of isometries of $S^3$ has the direct product form 
\begin{equation}
\label{h8}
SO(4,R) = SU^l(2) \times SU^r(2)/Z_2.
\end{equation}
Here $Z_2= \{e,-e\}$ is the group consisting of the unit $2\times 2$ 
matrix and its negative.

\section{Klein's  fundamental invariant of ${\cal H}_3$.}

F Klein \cite{KL}, 1884 in his monograph  - Vorlesungen \"uber das Ikosaeder- 
implements the E Galois 1847 theory of $A_5$.
He lets  the binary icosahedral group ${\cal H}_3 <SU(2)$
act by linear fractional transforms on two 
complex projective coordinates $(z_1,z_2),\; \zeta=(z_1/z_2)$ as 
\begin{eqnarray}
\label{h8a}
\zeta \rightarrow &&\zeta'=\frac{a\zeta-\overline{b}}{b\zeta+\overline{a}}\;,
\\ \nonumber
&&\left[
\begin{array}{ll}
a & b\\
-\overline{b} & \overline{a}
\end{array}
\right]=g_r \in SU(2,C), a\overline{a}+b\overline{b}=1.
\end{eqnarray}
Eq. \ref{h8} rewritten in terms of $(z_1,z_2)$ 
is exactly the right action in eq. \ref{h7} of $g_r$ from 
eq. \ref{h8} on $u \in S^3$ from eq. \ref{h6}.
Klein constructs   a ${\cal H}_3$-invariant complex polynomial,
\begin{equation}
\label{h9}
f_k(z_1,z_2):= (z_1z_2)\left[(z_1z_1)^5+11(z_1z_2)^5-(z_2z_2)^5\right],
\end{equation} 
of degree $12$ from the coordinates of the midpoints of the twelve 
dodecahedral faces. We shall see below that this invariant is in fact
a lowest degree harmonic polynomial on ${\cal P}$. Moreover we 
shall build from this particular invariant polynomial 
an invariant operator-valued polynomial which quantizes 
any harmonic polynomial on ${\cal P}$.

\section{The group of deck transformations.}

For the construction, we use a Coxeter group, H S M Coxeter \cite{CO}
pp. 187-212, which can be placed, eq. \ref{h12}, in between the continuous  group of
isometries of $SO(4,R)$ eq. \ref{h8} and the discrete group ${\cal H}_3$
of deck transformations which produces the dodecahedral tiling.
A Coxeter group is  finitely generated by involutive generators
and relations, coded in a Coxeter-Dynkin diagram. The relevant 
spherical Coxeter group with icosahedral subgroup
has the  Dynkin diagram, the four involutive generators and 
non-trivial relations 
\begin{eqnarray}
\label{h10}
&&\;\circ\;\frac{5}{}\;\circ\;\frac{3}{}\;\circ\; 
\frac{3}{}\;\circ\;:=
\\ \nonumber 
&&\langle  R_1,\,R_2,\,R_3,\,R_4\;|
\\ \nonumber
&&(R_1)^2=(R_2)^2=(R_3)^2=(R_4)^2
\\ \nonumber
&&=(R_1R_2)^5=(R_2R_3)^3=(R_3R_4)^3=I \rangle
\end{eqnarray}
Any pair of generators unlinked in the Dynkin diagram commutes.
A Coxeter group has a linear isometric representation 
by Weyl reflections (\cite{KR} pp. 3520-2), one for each generator.
The first three generators 
in eq. \ref{h10} form the icosahedral Coxeter group 
$\circ \frac{5}{}\circ \frac{3}{}\circ$ and 
generate from a 3-simplex in $S^3$ a dodecahedron.

This dodecahedron ${\cal P}$ is the prototile of eq. \ref{h3} 
under the group ${\rm deck}(\tilde{{\cal P}})$ 
of deck transformations.  
We find the following results \cite{KR}:

({\bf A1}): The group of deck transformations is a {\bf subgroup of the 
Coxeter group} eq. \ref{h10}.  

We construct a first Weber-Seifert gluing generator $C_1$ according to 
Fig. 1.

({\bf A2}): We  can express  explicitly the Weber-Seifert gluing generator $C_1$, lifted
to  $deck(\tilde{\cal P})$, as a product of
Coxeter group elements,
\begin{equation}
\label{h11}
C_1\;=\;R_4\;5_1^{-2}\;{\cal I}={\rm even},\; \{5_1,{\cal I}\}
\in  \circ \frac{5}{}\circ \frac{3}{}\circ.
\end{equation}
Here $5_1$ is a 5fold rotation around the vertical axis in Fig. 1, and ${\cal I}$
the inversion in the center of the dodecahedron.

By conjugation of $C_1$ with icosahedral rotations we get five more gluings, 
and so find\\ 
({\bf A3}): The group $deck(\tilde{\cal P})={\cal H}_3$ has the 
{\bf subgroup embedding}
\begin{equation}
\label{h12}
deck(\tilde{\cal P}) < S(\circ \frac{5}{}\circ \frac{3}{}\circ 
\frac{3}{}\circ) <SO(4,R).
\end{equation}
Here $S()$ denotes the unimodular restriction of the 
Coxeter group to $SO(4,R)$.

By computing from eq. \ref{h11} the action of $C_1\;=\;R_4\;5_1^{-2}{\cal I}$ eq. \ref{h11} on the 
complex coordinates eq. \ref{h6} of $S^3$ we find 
\begin{equation}
\label{h13}
C_1: (z_1,z_2)\rightarrow (z_1,z_2)\;
(-1)\left[
\begin{array}{ll}
\epsilon^{-2}&0\\
0&\epsilon^2
\end{array}
\right], \; \epsilon:=\exp(2\pi i/5).
\end{equation} 
The matrix in eq. \ref{h13} belongs to ${\cal H}_3$ (\cite{KR} p. 3524).

\begin{center}
\includegraphics{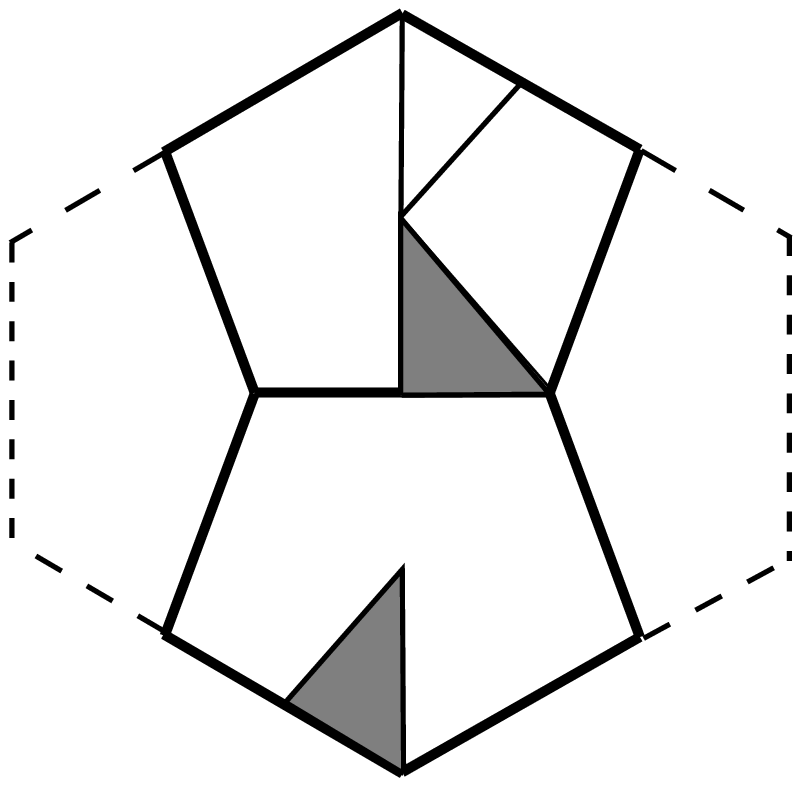} 
\end{center}

\textbf{Fig 1.}
{\bf Gluing prescription for ${\cal P}$ by Weber and Seifert}:
Two opposite pentagonal faces of the dodecahedral prototile 
in a projection along a 
2fold axis.
(1) The inversion ${\cal I}$ maps the shaded triangle from the bottom 
pentagon face to the white
triangle on the top face.
(2) The operation  $5_1^{-2}$ rotates the white triangle on the  top face 
to the shaded position. 
(3) The Weyl generator  
$R_4$, a reflection in the 
top pentagon face, glues the dodecahedron to its top face neighbour. 
\vspace{1cm}

({\bf A4}): The group $deck(\tilde{{\bf {\cal P}}})<SU^r(2,C)$ acts {\bf from the right} 
on $u \in S^3$, therefore 
commutes  with the continuous left action of $SU^l(2,C)$.

({\bf A5}): The group $deck(\tilde{\cal{P}})$ is {\bf normal} 
in the unimodular restriction of the Coxeter group 
and forms with the icosahedral
rotation group $S(\circ \frac{5}{}\circ \frac{3}{}\circ)$ 
the semidirect product:
\begin{equation}
\label{h14}
S(\circ \frac{5}{}\circ \frac{3}{}\circ 
\frac{3}{}\circ)
= deck(\tilde{P}) \times_s  S(\circ \frac{5}{}\circ \frac{3}{}\circ). 
\end{equation}

\section{From group actions to representation spaces:\\ 
Harmonic polynomials.}

Suppose  we would adopt the simply connected 3-sphere $S^3$ as the space part of the
cosmological model. Then we need the harmonic polynomials 
on $S^3$. 

Harmonic polynomials $P$ on $S^3$, homogeneous of  degree $\lambda$,
obey  $\Delta P=0$ where $\Delta$ is the Laplacian on $S^3$.
We find: 

The harmonic polynomials on $S^3$ in the complex coordinates 
eq. \ref{h6}
are identical with 
Wigner's standard  
irrep matrices of $SU(2)$, \cite{ED} pp. 53-67, taken as polynomials of degree 
$\lambda=2j$ in the elements of the matrix eq. \ref{h6} $u \in SU(2)$:
\begin{eqnarray}
\label{h15}
&&P^{\lambda}(z_1,z_2,\overline{z}_1,\overline{z}_2)
=D^j_{m,m'}(z_1,z_2,\overline{z}_1,\overline{z}_2)=
\left[\frac{(j+m')!(j-m')!}{(j+m)!(j-m)!}\right]^{1/2}
\\ \nonumber
&&\sum_{\sigma}
\frac{(j+m)!(j-m)!}{(j+m'-\sigma)!(m-m'+\sigma)! (j-m-\sigma)!\sigma!}
z_1^{j+m'-\sigma}(-\overline{z}_2)^{m-m'+\sigma}
(\overline{z}_1)^{j-m-\sigma}z_2^{\sigma},
\\ \nonumber 
&&-j \leq (m,m') \leq j,\; 
\end{eqnarray}
For given degree $\lambda=2j$, there are $(2j+1)^2$ orthogonal
harmonic polynomials.

{\em Proof}: (i) The degree of 
$D^j_{m,m'}(z_1,z_2,\overline{z}_1,\overline{z}_2)$ is $\lambda=2j$. 
(ii) For $m=j$, $D^j$ is analytic in $(z_1,z_2)$, hence $\Delta D^j_{j,m'}=0$.
(iii) Since $m$ can be lowered by the operator $L^l_-$ commuting with $\Delta$,\\
\begin{equation}
\label{h15l}
\Delta D^j_{m,m'} \sim \Delta (L^l_-)^{j-m}D^j_{j,m'}
=(L^l_-)^{j-m} \Delta D^j_{j,m'} =0.
\end{equation}

If the space part of the cosmology is modelled by $S^3$,
the basis eq. \ref{h15} should allow to expand the CMB fluctuations.
In contrast, the harmonic polynomials for the Poincar\'{e} model ${\cal P}$
must be a subset
of those in eq. \ref{h15}, since they must repeat their functional
values on all copies of the spherical Poincar\'{e} dodecahedron which tile
$S^3$.

In terms of group representations, the problem of finding the harmonic polynomials on the Poincare manifold
${\cal P}$ 
can be formulated as follows:
From all the harmonic polynomials eq. \ref{h15} on $S^3$, 
we must select the subset
which belongs 
to the identity irrep $D^{\alpha_0}\equiv 1$ of 
${\cal H}_3 \in SU^r(2)$.

For fixed irrep $j$ of $SU^r(2)$ with character $\chi^j$
we can compute the multiplicity $m(j,\alpha_0)$  
of ${\cal H}_3$-invariant polynomials  
from a scalar product of characters, 
\begin{equation}
\label{h15a}
m(j,\alpha_0) :=\frac{1}{|{\cal H}_3|}\sum_{g\in {\cal H}_3}
\chi^j(g) \overline{\chi}^{\alpha_0}(g),\; \chi^{\alpha_0}(g)=1,
\end{equation} 
with the following result:

The multiplicity $m(j,\alpha_0)$  of invariant harmonic polynomials
is zero for degree $2j={\rm odd}$. For $2j={\rm even}$ 
it is given by
\begin{eqnarray}
\label{h15b}
&&(i)\;\; {\rm the}\; {\rm starting}\;\; 
{\rm  values}\; m(j,\alpha_0)=1\;\; {\rm for}\;\; j < 30:
\\ \nonumber
&& j=0,6,10,12,15,16,18,20,21,22,24,25,26,27,28,\; m=0\;\; {\rm otherwise},
\\ \nonumber
&&(ii)\;\; {\rm the}\; {\rm recursion}\;\;{\rm relation}\;\; {\rm from}\;\; 
{\rm the}\; {\rm characters}:
\\ \nonumber
&&m(j+30,\alpha_0)=m(j,\alpha_0)+1.
\end{eqnarray}
The multiplicity eq. \ref{h15b} characterizes the subduction
$SU^r(2) > {\cal H}_3$. In addition, since ${\cal H}_3$ and
$SU^l(2)$ commute, there is for fixed $j$ an additional multiplicative
degeneracy $(2j+1)$ of harmonic polynomials, see {\bf B4} below.

From ({\em ii}) we conclude that the relative  fraction up to 
integer $j$
of harmonic polynomials for  $({\cal P}/S^3)$ on average is
$\sum_0^j m(j',\alpha_0)/(\sum_0^j(2j'+1)) \sim 1/30$.

{\bf There are strong selection rules and a low mode suppression} 
for $2j <12$ of harmonic polynomials on
${\cal P}$ versus those on its universal cover $S^3$. Weeks \cite{WE} 
phrases this as the 
{\em Mystery of the missing fluctuations}.
The Poincar\'{e} model would be verified if the expansion of the CMB fluctuations
in terms of the harmonic polynomials eq. \ref{h15} would display the selectivity stated in eq. \ref{h15b}.

\section{Group/subgroup subduction of  irreps by\\ a generalized Casimir operator.}

We proceed to the explicit determination of the invariant polynomials.
For this purpose we first extend the problem and find the full
group/subgroup subduction in $SU(2)>{\cal H}_3$ for all irreps of ${\cal H}_3$, 
and then from these select the identity irrep $D^{\alpha_0}$. 

We follow the procedure from 
V Bargmann and M Moshinsky \cite{BA} 1960,
exemplified by them for $SU(3,C)>SO(3,R)$:\\
A  {\bf generalized Casimir operator} $\Omega$ 
determines the irrep subduction 
$G>H$  iff (i,ii,iii) hold:\\
(i) $\Omega$ is from the enveloping Lie algebra $Env(l_G)$ and so preserves
irrep spaces under $G$ ,\\
(ii) $\Omega$ is invariant under $H$ but not under $G$,\\
(iii) $\Omega$ is non-degenerate.\\
Part (iii) of this definition excludes the Casimir or projection
operators of $H$ constructed in $Env(l_G)$, since they cannot distinguish between
repeated irreps!

As tools we determine the  right action generators of\\
$SU^r(2)$ acting on $S^3$ eq. \ref{h7}:

The right Lie generators of $SU^r(2)$ from eq. \ref{h7} act on functions of $(z_1,z_2,\overline{z}_1,\overline{z}_2)$ 
as first order differential  operators:
\begin{eqnarray}
\label{h16}
&& L_+:= L^r_1+iL^r_2=\left[z_1\partial_{z_2}
-\overline{z}_2\partial_{\overline{z}_1}\right],
\\ \nonumber
&& L_-:= L^r_1-iL^r_2=\left[z_2\partial_{z_1}
-\overline{z}_1\partial_{\overline{z}_2}\right],
\\ \nonumber
&& L_3:= L^r_3
 =(1/2)\left[z_1\partial_{z_1}-z_2\partial_{z_2}
-\overline{z}_1\partial_{\overline{z}_1}
+\overline{z}_2\partial_{\overline{z}_2}\right],
\\ \nonumber
&& \left[L_3,L_{\pm}\right]= \pm L_{\pm},\;
\left[L_+,L_-\right]= 2L_3.
\end{eqnarray}
The left Lie generators of $SU^l(2)$ from eq. \ref{h7} look similar but commute 
with all the right Lie generators (\cite{KR} pp. 3525-6).

The next tool is  Klein's homomorphism 
$SU^r(2) \rightarrow SO(3,R)$:

Under the right action
\begin{equation}
\label{h17} 
(z_1,z_2) \rightarrow (z_1,z_2)\;g_r,\; g_r \in SU(2),
\end{equation}
the vector 
\begin{eqnarray}
\label{h17a}
&&((x+iy)/\sqrt{2},z,(x-iy)/\sqrt{2})
\\ \nonumber 
&&\equiv (2z_1\overline{z}_2, z_1\overline{z}_1-z_2\overline{z}_2,
2\overline{z}_1z_2),  
\end{eqnarray}
and under conjugation 
\begin{equation}
\label{h18}
(L_+, L_3, L_-) \rightarrow U_{g_r}\;(L_+, L_3, L_-)\;U_{g_r}^{-1}
\end{equation}
the right action generators eq. \ref{h16} transform linearly {\bf with  
the same representation} $D^1(g_r) \in SO(3,R)$!

In the following steps we apply the  noncommutative geometry
of operators from the enveloping algebra $Env(su(2))$, similar to the Penrose length quantization in spin networks
and to quantum gravity.

\section{Lie-algebraic results on\\
 harmonic polynomials of ${\cal P}$.}

To find harmonic polynomials we proceed \cite{KR} as follows :\\
({\bf B1}): Construct a  ${\cal H}_3$-invariant polynomial 
\begin{eqnarray}
\label{h19}
{\cal K}'&=&P(2z_1\overline{z}_2, z_1\overline{z}_1-z_2\overline{z}_2,
2\overline{z}_1z_2)
\\ \nonumber 
&=& P((x+iy)/\sqrt{2},z,(x-iy)/\sqrt{2}).
\end{eqnarray}
Klein's invariant polynomial eq. \ref{h9} cannot be written in terms of the 
vector
components eq. \ref{h17a} which then would allow to pass to the generators 
$(L_+, L_3, L_-)$. Fortunately we can generate other invariant polynomials
by applying the left lowering
generator $L^l_-$ from $SU^l(2)$ to 
Klein's fundamental invariant eq. \ref{h9}.
Applying the power $(L^l_-)^6$ to Klein's invariant one obtains 
(\cite{KR} pp. 3526-8)
a polynomial
in which the vector components of the homomorphism eq. \ref{h17} 
can be substituted:
\begin{eqnarray}
\label{h20}
&&{\cal K}' \sim  (L^l_-)^6f_k(z_1,z_2)\\
\nonumber
&& \sim  (x+iy)^5z + z(x-iy)^5 + P_2(r^2,z) :=P.
\end{eqnarray}\\

({\bf B2}): Use the Klein homomorphism eq. \ref{h17a}, substitute\\ 
${\cal K}'(x+iy,z,x-iy) \rightarrow {\cal K}(L_+,L_3,L_-)$.\\
Since now we are dealing  with homogeneous polynomials of degree 
$n$,
we run into the {\em noncommutativity problem} of operator-valued 
polynomials in $(L_+,L_3,L_-)$: 
To assure the same transformation of the  polynomial 
operator ${\cal K}$ as for ${\cal K}'$,  
we must apply the operator of symmetrization defined by 
\begin{equation}
\label{h20a}
{\rm Sym}\; P(A_1,\ldots, A_n):= 
\frac{1}{n!} \sum_{(i_1,\ldots,i_n) \in S_n} P(A_{i_1},\ldots,A_{i_n}).
\end{equation}
In this way we get from eqs. \ref{h20}, \ref{h20a} the 
$H_3$-{\bf invariant hermitian generalized Casimir operator}
\begin{eqnarray}
\label{h20b} 
{\cal K}&:=& {\rm Sym} \left[{\cal K}(L_+,L_3,L_-)\right] \in Env(su^r(2)).
\\ \nonumber
    &=&  {\rm Sym}\left[
(L_+)^5L_3+ L_3(L_-)^5 +  P_2(L^2,L_3)\right]
\end{eqnarray}
By construction ${\cal K}$ 
commutes with the Casimir operator $\Lambda^2$
of $SO(4,R)$.
Note that ${\rm Sym}$ has 6!=720 terms. The symmetrization  eq. \ref{h20b}
is performed
in the appendix
of \cite{KR}.

({\bf B3}): Quantize the spherical harmonics $P^{2j}$ 
by diagonalizing the right action of ${\cal K}$. The eigenspaces are characterized
by irreps $D^{\alpha}$ of ${\cal H}_3$.

({\bf B4}): ${\cal K}$ commutes with $SU^l(2,C)$, so the  
degeneracy of any eigenvalue $\kappa$ of ${\cal K}$ is $(2j+1)dim(\alpha)$,
and the harmonic polynomials on $S^3$ are now characterized by 
\begin{eqnarray}
\label{h21}
&&P= P^{2j}_{m,\kappa}(z_1,z_2,\overline{z}_1,\overline{z}_2):
\\ \nonumber 
&&\Delta P^{2j}_{m,\kappa}=0,\; 
(x \cdot \nabla) P^{2j}_{m,\kappa}=(2j)P^{2j}_{m,\kappa},
\\ \nonumber 
&&\Lambda^2P^{2j}_{m,\kappa}=4j(j+1)P^{2j}_{m,\kappa},
\\ \nonumber
&&{\cal K}P^{2j}_{m,\kappa}=\kappa P^{2j}_{m,\kappa},\; 
L^l_3P^{2j}_{m,\kappa}=mP^{2j}_{m,\kappa}.
\end{eqnarray}

({\bf B5}): {\bf Harmonic polynomials on ${\cal P}$}  must belong to
the {\bf identity irrep} $\alpha_0$ of ${\cal H}_3$.

({\bf B6}): The {\bf Spectrum} of ${\cal K}$ (\cite{KR} pp. 3530-1): 
${\cal K}$ by eq. \ref{h20b} acts in linear subspaces 
${\cal L}^{\mu}: m\equiv \mu\;{\rm mod}\; 5$
and in these subspaces  is tridiagonal. By general reasons 
given in \cite{BA}, it follows from tridiagonality that
the operator ${\cal K}$ in any linear subspace  ${\cal L}^{\mu}$  
is {\bf non-degenerate} (no repeated eigenvalues).
Moreover it is shown in (\cite{KR} p.3531) that the identity irrep $D^{\alpha_0}$ of
${\cal H}_3$ can occur only in ${\cal L}^0$.

The hermiticity of ${\cal K}$ 
yields the orthogonality of the eigenstates.

The properties {\bf B1 ... B6} taken together show that the operator ${\cal K}$
by its eigenvalues and eigenstates completely characterizes 
the harmonic polynomials on the dodecahedral Poincar\'{e}
manifold ${\cal P}$. The unique characterization by ${\cal K}$
extends beyond the identity representation to any irrep 
$D^{\alpha}$ of ${\cal H}_3$. For the Poincar\'{e} model we
select only the polynomials belonging to the identity irrep
${\alpha}_0$ of ${\cal H}_3$.

For the full analysis of the subduction and detailed results
for degrees $2j=0,\ldots, 12$ we refer to \cite{KR}.
As an example for the diagonalization of ${\cal K}$ we  give case $j=6$.
The values of $\mu$ and corresponding values of $m$ are given  by
\begin{equation}
\label{h22}
\begin{array}{lll}
j=6:&\mu & m \\
    &0   & -5,0,5\\
    &1   & -4,1,6\\
    &2   & -3,2\\
    &3   & -2,3\\
    &4   & -6,-1,4\\
\end{array}
\end{equation}
From  {\bf B6}, we expect harmonic polynomials only in ${\cal L}^0$
of dimension 3. Evaluation of the operator ${\cal K}$ in this subspace
gives its matrix form and eigenvalues,

\begin{eqnarray}
{\cal K}\;V^{\mu}&=&V^{\mu}\;{\cal K}^{\mu, diag} 
\\ \nonumber
 j&=&6,\; \mu=0,\; m=(-5,0,5):
\\ \nonumber
\kappa &=& (-51975, -\frac{51975}{2}, 
\frac{14175}{2}).
\\ \nonumber
{\cal K}&=&
\left[
\begin{array}{rrr}
-\frac{51975}{2}&\frac{4725\sqrt{77}}{2}&0\\
\frac{4725\sqrt{77}}{2}&-18900&-\frac{4725\sqrt{77}}{2}\\
0&-\frac{4725\sqrt{77}}{2}& -\frac{51975}{2}
\end{array}
\right],
\\ \nonumber
V^{0}&=&\left[
\begin{array}{rrr}
-\sqrt{\frac{7}{25}}&\sqrt{\frac{1}{2}}&-\sqrt{\frac{11}{50}}\\
\sqrt{\frac{11}{25}}&0&-2\sqrt{\frac{7}{50}}\\
\sqrt{\frac{7}{25}}&\sqrt{\frac{1}{2}}&\sqrt{\frac{11}{50}}  
\end{array}
\right],
\\ \nonumber
{\cal K}^{0, diag}&=&\left[
\begin{array}{rrr}
-51975&&\\
&-\frac{51975}{2}&\\
&&\frac{14175}{2}
\end{array}
\right].
\end{eqnarray}

The first eigenstate with  
$\kappa= -51975$ is a harmonic polynomial and 
up to a constant factor turns out to be  Klein's $f_k(z_1,z_2)$ 
eq. \ref{h9}! Twelve more orthogonal companions with the same eigenvalue 
$\kappa$ are obtained by applying powers of the lowering operator $L^l_-$.

The other two eigenstates are not harmonic polynomials of ${\cal P}$ 
for the following reason:
It turns  out (\cite{KR} p. 3534) that they have companions with the same 
eigenvalue $\kappa$, 
but in subspaces 
${\cal L}^{\mu}, \mu \neq 0$. Together with these they span 
irreps $D^{\alpha}, \alpha \neq \alpha_0$ of ${\cal H}_3$ and therefore are not harmonic polynomials of
${\cal P}$.

The explicit diagonalization of ${\cal K}$ for degrees 
$2j \leq 12$ is given in (\cite{KR} pp. 3532-6). There is no problem in going
on to any higher degree harmonic polynomials as eigenstates of ${\cal K}$.
Since these harmonic polynomials are orthogonal, the expansion coefficients
of the observed CMB fluctuations in terms of harmonic polynomials 
are given by the scalar products between the observed fluctuations and these
polynomials. The strict validity of the Poincar\'{e} manifold as a model 
for the
space-part of the cosmos would imply that all scalar products beween the
fluctuations and harmonic polynomials not belonging to the identity
irrep $D^{\alpha_0}$ of ${\cal H}_3$ must vanish.

\section{Conclusion.}

$\bullet$ The subduction $SO(4,R)>{\cal H}_3$ for {\bf any irrep} of 
${\cal H}_3$ is explicitly resolved by   
the operator ${\cal K}$. Harmonic polynomials on ${\cal P}$ become 
(non-degenerate) 
eigenstates of ${\cal K}$ eq. \ref{h20b}!

$\bullet$ For degree $2j=12$, only Felix Klein's invariant harmonic polynomial 
$f_k$ eq. \ref{h9} plus twelve orthogonal companions belong to the 
non-degenerate eigenvalue  $\kappa= -51975$ of ${\cal K}$.
All other eigenstates have degenerate companions in subspaces
${\cal L}^{\mu'},\; \mu' \neq 0$ belonging to irreps 
$\alpha \neq \alpha_0$.\\
$\bullet$ There is an additional controlled degeneracy $(2j+1)$ from invariance under
$SU^l(2,C)$.

$\bullet$ If 3-space has the topology of ${\cal P}$, we can expand 
the temperature fluctuations of CMB exclusively in 
invariant eigenmodes of ${\cal K}$!

$\bullet$ A similar analysis can be done for the topological 3-manifolds $S^3/T^*,S^3/O^*$.
with $T^*,O^*$ the binary tetrahedral, octahedral group. All these 
3-manifolds share $S^3$ as their universal cover. 

$\bullet$ What about
hyperbolic 3-manifolds like the Weber-Seifert manifold \cite{KR3}? Here the universal cover is the hyperbolic
space of dimension 3. All hyperbolic 3-manifolds have 
negative curvature $\kappa=-1$.

\end{document}